\documentclass
[aps,twocolumn,prb,showpacs,preprintnumbers,amsmath,amssymb]{revtex4}
\usepackage{natbib}
\usepackage[dvips]{graphicx}
\usepackage{dcolumn}
\usepackage{bm}
\usepackage{times}
\usepackage{mathptmx}
\usepackage[usenames]{color}

\begin{document}
\draft

\title{Spin-state transition and phase separation in multi-orbital Hubbard model}
\author{Ryo~Suzuki$^{1 \dagger}$, Tsutomu~Watanabe$^{2 \ast}$, and Sumio~Ishihara$^1$} 
\affiliation{
$^1$Department of Physics, Tohoku University, Sendai 980-8578, Japan\\
$^2$Institute of Multidisciplinary Research for Advanced Materials, Tohoku University, Sendai 980-8577, Japan}
\date{\today}

\begin{abstract}
We study spin-state transition and phase separation involving this transition
based on the milti-orbital Hubbard model. 
Multiple spin states are realized by changing the energy separation between the two orbitals and the on-site Hund coupling. 
By utilizing the variational Monte-Carlo simulation, we analyze the electronic and magnetic structures in hole doped and undoped states. 
Electronic phase separation occurs between the low-spin band insulating state and the high-spin ferromagnetic metallic one. 
Difference of the band widths in the two orbitals is of prime importance for the spin-state transition and the phase separation.
\end{abstract}

\pacs{75.25.+z, 71.70.-d, 71.30.+h} 
%75.25.+z, Spin arrangements in magnetically ordered materials 
%71.70.-d, Level splitting and interactions
%71.30.+h Metal--insulator transitions and other electronic transitions

\maketitle
\narrowtext

Novel electric and magnetic phenomena observed in correlated electron systems are responsible for competition and cooperation between multi-electronic phases with delicate energy balance. 
These are owing to the internal degrees of freedom of electrons, i.e. spin, charge and orbital, under strong electron correlation, and their coupling with crystal lattice.~\cite{book,imada} 
In some transition-metal ions, there is an additional degree of freedom, 
termed the spin-state degree of freedom, i.e. 
multiple spin states due to the different electron configurations in a single ion.  
One prototypical example is the perovskite cobaltites $R_{1-x}A_x$CoO$_3$ ($R$: rare earth ion, $A$: alkaline earth ion)  where transitions between the multiple spin states occur by changing carrier concentration, temperature and so on. 
In Co$^{3+}$ with the $d^6$ configuration, there are three possible spin states, the high-spin (HS) state
$(e_g^2t_{2g}^4)$ with an amplitude of $S=2$, the intermediate-spin (IS) one $(e_{g}^1t_{2g}^5)$ with $S=1$, and the low-spin (LS) one $(t_{2g}^6)$ with $S=0$. 

Several magnetic, electric and transport measurements have been carried out in the insulating and metallic cobaltites. 
It is known that LaCoO$_3$ is a non-magnetic LS band-insulator (BI) at low temperatures, 
although there is still controversy in the spin-state transition and the IS state at finite temperature.~\cite{yamaguchi,korotin,haverkort,noguchi,kobayashi2}  
In high hole doping region of $x>0.3-0.4$ in La$_{1-x}$Sr$_x$CoO$_3$, the ferromagnetic (FM) metallic state was experimentally confirmed. 
In the lightly hole doped region between the two, 
a number of inhomogeneous features in magnetic, electric and lattice structures have been reported experimentally.   
Spatial segregation of hole-rich FM regions and hole-poor insulating ones have been suggested by the neutron diffraction, the electron microscopy, NMR and so on.~\cite{itoh,caciuffo,kuhns,ghoshray} 
Magnetic/non-magnetic clusters have been found by the small-angle and inelastic neutron scattering experiments.~\cite{wu,phelan1,phelan2} 
It is widely believed that the observed giant magneto-resistance effect in the lightly doped region 
results from the electronic and magnetic inhomogeneity.~\cite{wu} 

Electronic phase separation (PS) phenomena in transition-metal compounds have been studied extensively and intensively, in particular, in the high Tc superconducting cuprates and the colossal magnetoresistive manganites.~\cite{nagaev,dagotto,okamotos,kugel} In these materials, the long-range spin/orbital orders in the Mott insulating phases and their melting by carrier doping are of essence in the electronic PS. 
The exchange energy for the localized spins/orbitals and the kinetic one for the itinerant electrons 
are gained in spatially separate regions. 
On the other hand, in the present case, the non-magnetic band insulator is realized in the insulating phase, 
and the spin-state transition is brought about by carrier doping. 
Thus, the present phenomena belong to a new class of the electronic PS in correlated system, 
although only a little theoretical studies have been done until now. 
In this paper, we address the issues of the spin-state transition and the PS associated with this transition by analyzing the multi-orbital Hubbard model. We examine the electronic structures in hole doped and undoped systems by utilizing the variational Monte-Carlo (VMC) method. 
We find that, between the non-magnetic BI and the HS FM metal, 
the electronic PS is realized. 
We claim that the different band widths play an essential role in the present electronic PS.   

We set up a minimal model, the two-orbital Hubbard model,~\cite{werner,sano,kobayashi,kubo} where the spin-state degrees of freedom and a transition between them are able to be examined. 
In each site in a crystal lattice, we introduce two orbitals, termed A and B, 
which represent one of the $e_g$ and $t_{2g}$ orbitals, respectively. 
Anisotropic shape of the orbital wave function is not concerned.  
An energy difference between the two orbitals is denoted by $\Delta \equiv \varepsilon_A-\varepsilon_B >0$  
where $\varepsilon_A$ ($\varepsilon_B$) is the level energy for A(B). 
When the electron number per site is two, the lowest two electronic states in a single site are
$| B^2 \rangle$ and $|A^1 B^1 \rangle$ with triplet spin state which are termed the LS and HS states in the present model, respectively. 
The explicit form of the model Hamiltonian is given by 
\begin{eqnarray}
{\cal H}&=&\Delta \sum_{i \sigma} c_{i A \sigma}^\dagger c_{i A \sigma}
- \sum_{\langle ij \rangle \gamma \sigma} 
t_\gamma
\left ( c_{i \gamma \sigma}^\dagger c_{j \gamma \sigma} +H.c. \right ) 
\nonumber \\
&+&U\sum_{i \gamma} n_{i \gamma \uparrow} n_{i \gamma \downarrow}
+U'\sum_{i \sigma \sigma'} n_{i A \sigma} n_{i B \sigma'}
\nonumber \\
&-&J\sum_{i \sigma \sigma'}c_{i A \sigma}^\dagger c_{i B \sigma} c_{i B \sigma'}^\dagger c_{i A \sigma'}
-J'\sum_{i \gamma} c_{i \gamma \uparrow}^\dagger c_{i {\bar \gamma} \uparrow} c_{i \gamma \downarrow}^\dagger c_{i {\bar \gamma} \downarrow} , 
\label{eq:ham}
\end{eqnarray}
where $c_{i \gamma \sigma}$ is the annihilation operator of an electron at site $i$ with orbital $\gamma(={A, B})$ and spin $\sigma(=\uparrow , \downarrow)$, and 
$n_{i \gamma \sigma}\equiv c_{i \gamma \sigma}^\dagger c_{i \gamma \sigma}$ is the number operator. 
A subscript ${\bar \gamma}$ takes $ A(B)$, when $\gamma$ is $ B(A)$. 
We assume that the transfer integral is diagonal with respect to the orbitals and 
$|t_A| > |t_B|$, both of which are justified in perovskite cobaltites. 
In most of the numerical calculations, a relation $t_B/t_A=1/4$ is chosen. 
As the intra-site electron interactions, 
we introduce the intra- and inter-orbital Coulomb interactions, $U$ and $U'$, respectively, the Hund coupling $J$ and the pair-hopping $J'$. 
The relations $U=U'+2J$ and $J=J'$ satisfied in an isolated ion are assumed. 
In addition, we introduce the relation $U=4J$ in the numerical calculation. 

We adopt the VMC method where the electron correlation is treated in an unbiased manner 
and simulations in a large cluster size are possible. 
For simplicity and a limitation in the computer resource, we introduce two-dimensional square lattices 
with a system size of $N \equiv L^2$ ($L \leq 6$) and the periodic and anti-periodic boundary conditions. 
The number of electron is $N_e$, and the hole concentration per site measured from $N_e=2N$ is denoted as $x \equiv (2N-N_e)/N$. The variational wave function is given as a product form of 
$\Psi=G |\Phi \rangle$ where $G$ is the correlation factor and $|\Phi \rangle$ is the one-body wave function. 
The two types of the wave function are considered in $|\Phi \rangle$: 
the Slater determinant obtained by the second term in Eq.~(\ref{eq:ham}), and that for the HS antiferromagnetic (AFM) order given by applying the Hartree-Fock approximation to the third term in Eq.~(\ref{eq:ham}).  
In the latter, the AFM order parameter is treated as a variational parameter.  
We assume the Gutzwiller-type correlation factor $\Pi_{i l}(1-\xi_l {\cal P}_{il})$ where $l$ indicates the local electron configurations, ${\cal P}_{i l}$ is the projection operator at site $i$ for the configuration $l$, and $\xi_l$ is the variational parameter. 
Here we introduce the 10 variational parameters for the 10 inequivalent electron configurations in a single site.~\cite{variation} 
The fixed-sampling method is used to optimize the variational parameters.~\cite{umrigar} 
In addition to the standard VMC method, we improve the variational wave function by estimating analytically the weights 
for the configurations which are sampled by the MC simulations. 
This method is valid for the LS state and reduces the CPU time by more than one order.    
In most of the calculations, $10^4-10^5$ MC samples are adopted for measurements.
%and a numerical error of the energy expectation is of the order of $10^{-3}t_A$. 

We start from the case at $x=0$ where 
the average electron number per site is two. 
The electronic states obtained by the simulation are monitored 
by the total spin amplitude defined by ${\bf S}^2=(1/N) \sum_i 
\langle {\bf S}_i^2 \rangle$ where ${\bf S}_i=\sum_\gamma {\bf S}_{ i \gamma} =(1/2)
\sum_{s s'\gamma} c_{i \gamma s}^\dagger {\bf \sigma}_{s s'} c_{i \gamma s'}$ is the spin operator 
with the Pauli matrices $\bf \sigma$, 
the spin correlation function 
$S_\gamma({\bf q})=(4/N) \sum_{ij } 
e^{i {\bf q} \cdot( {\bf r}_i-{\bf r}_j)} \langle  S^z_{i \gamma} S^z_{j \gamma} \rangle$, 
and the momentum-distribution function 
$n_\gamma ({\bf k})=(1/2)\sum_{\sigma} \langle c_{{\bf k} \gamma \sigma}^\dagger c_{{\bf k} \gamma \sigma} \rangle$ 
where $c_{{\bf k} \gamma \sigma}$ is the Fourier transform of $c_{i \gamma \sigma}$. 
Size dependences of ${\bf S}^2$ and $S_\gamma({\bf q})$ in $L=4-8$ are within a few percent. 
We obtain the three phases, the HS Mott insulator (MI), the LS BI and the metallic (ML) phase.   
In the HS-MI phase, ${\bf S}^2$ is about 1.6 being about 80$\%$ of the maximum value for $S=1$. 
A sharp peak in $S_\gamma ({\bf q})$ at ${\bf q}=(\pi, \pi)$ and no discontinuity in $n_{\gamma}({\bf k})$ 
imply that this is the AFM MI. 
In the LS-BI phase, $n_{ A}({\bf k})$ $[n_{B}({\bf k})]$ is almost zero (one) in all momenta, and 
${\bf S}^2 \simeq 0$. 
In the ML phase, discontinuous jumps are observed in both $n_{ A}({\bf k})$ and $n_{ B}({\bf k})$. 
The electron (hole) fermi surface is located around ${\bf k}=(0,0)$ $[(\pi, \pi)]$ in the A (B) band; 
this is a semi metal. 
A value of ${\bf S}^2$ is about 0.3, and no remarkable structure is seen in $S_\gamma ({\bf q})$. 

\begin{figure}[t]
\begin{center}
\includegraphics[width=1.0\columnwidth,clip]{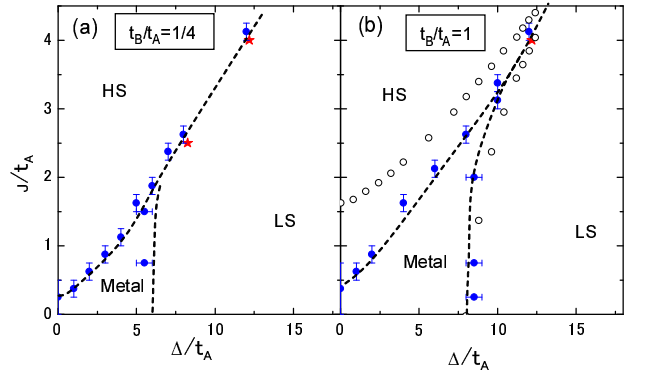}
\caption{(color online) 
Phase diagrams at $x=0$. A ratio of the electron transfers is taken to be  $t_{B}/t_{A}=1/4$ in (a) and $t_{B}/t_{A}=1$ in (b). 
In (b), filled squares and open circles are for the results obtained by the VMC method and the previous DMFT one in Ref.~\protect{\onlinecite{werner}}, respectively.  
Broken curves are guides for eyes. 
Stars represent the parameters where the carrier dopings are examined. 
}
\label{fig:fig1}
\end{center}
\end{figure} 
The phase diagram at $x=0$ is presented in Fig.~\ref{fig:fig1}. 
The error bars imply the upper and lower bounds of the phase boundary, and symbols are plotted at the middle of the bars.  
In the region of large $\Delta$ ($J$), the LS-BI (HS-MI) phase is realized, and between the two with small $\Delta$ and $J$, the ML phase appears. 
To compare the present results with the previous ones calculated by the dynamical-mean field theory (DMFT),~\cite{werner} we present, in Fig.~\ref{fig:fig1}(b), the phase diagram where the two transfer integrals are chosen to be equal, i.e. $t_{B}/t_{A}=1$.
%The three sets of the data are shown: the DMFT results, the VMC ones, and the VMC ones where 
%the long-range AFM order is not taken into account in the one-body part of the variational wave function. 
%When we compare the results with $t_{A}/t_{B}=1/4$, the metallic phase is stabilized in a wide parameter region. 
%This is due to the increasing of the kinetic energy gain in the metallic phase in the $t_{\rm A}/t_{\rm B}=1$ case. 
%Focus on the difference between the results by the VMC and DMFT methods. 
Although the global features in the phase diagrams are the same with each other, 
the HS-MI phase obtained by the VMC method appears in a broader parameter region than that in DMFT, 
in particular, near the boundary of the HS-MI and ML phases. 
This is because the AFM long-range order in the HS-MI phase is treated properly 
in the VMC method.
We have confirmed that the phase boundaries obtained by the VMC method where the AFM order is not considered 
almost reproduce the DMFT results. 

\begin{figure}[t]
\begin{center}
\includegraphics[width=0.8\columnwidth,clip]{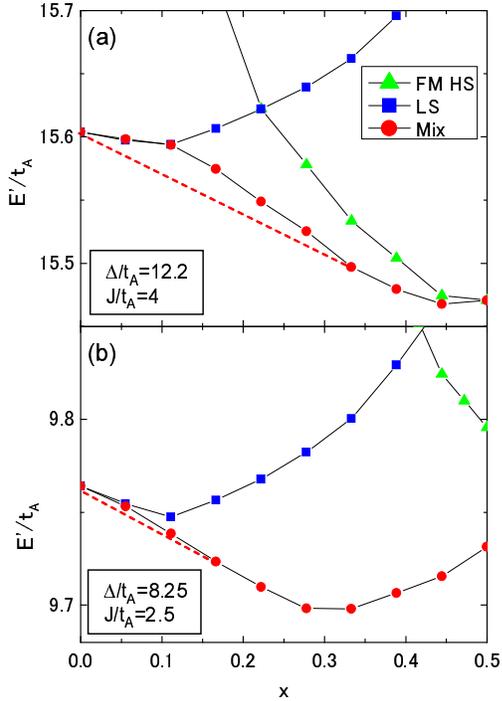}
\caption{(color online) 
Hole concentration dependences of the energy expectations for several states at 
$(\Delta/t_{\rm A}, J/t_{\rm A})=(12.2, 4)$ in (a), 
and those at $(\Delta/t_{\rm A}, J/t_{\rm A})=(8.25, 2.5)$ in (b). 
Broken lines are given by the Maxwell's construction. 
A ratio of the electron transfers is taken to be $t_{B}/t_{ A}=1/4$. 
A constant parameter $C$ in the definition of $E'$ is taken to be 8.2 in (a) and 5.25 in (b). 
}
\label{fig:fig2}
\end{center}
\end{figure} 
Now we show the results at finite $x$.
Holes are introduced into the LS-BI phase near the phase boundary  
with the parameter values of $(\Delta/t_{\rm A}, J/t_{\rm A})=(12.2, 4)$ and $(8.25, 2.5)$ [see Fig.~\ref{fig:fig1}]. 
By changing the initial conditions in the VMC simulation, we obtain the following four states: 
i) the LS-ML state where $n_{\rm A}({\bf k})$ is almost zero in all ${\bf k}$, and the fermi surface is located in the B band around ${\bf k}=(\pi, \pi)$, 
ii) the FM HS-ML state where $n_{\rm B}({\bf k})$ is about $1/2$ in all ${\bf k}$, the fermi surface is in the A band, and $S_\gamma({\bf q})$ has a sharp peak at ${\bf q}=(0,0)$, 
iii) the AFM HS-ML state where the fermi surface exists in the A band around ${\bf k}=(\pi, 0)$, and $S_\gamma({\bf q})$ has  a peak at ${\bf q}=(\pi, \pi)$, and 
iv) the mixed state where the wave function is a linear-combination of the LS-ML and FM HS-ML states. 

\begin{figure}[t]
\begin{center}
\includegraphics[width=0.8\columnwidth,clip]{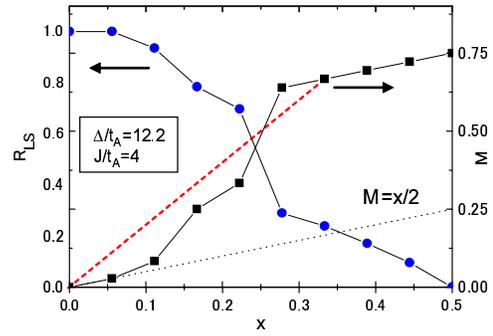}
\caption{(color online) 
A ratio of the LS sites to the LS and HS ones in the mixed state, $R_{LS}$, and magnetization $M(x)$ as functions of the hole concentration $x$. 
A broken line connecting data at $M(x=0)$ and $M(x=0.33)$ is drawn by the Maxwell's rule. 
For comparison, we plot a $M(x)=x/2$ curve which is expected from the hole doping in the LS-BI phase.
Parameters are chosen to be 
$(\Delta/t_{\rm A}, J/t_{\rm A})=(12.2, 4)$ and $t_{B}/t_{A}=1/4$.}
\label{fig:fig3}
\end{center}
\end{figure}

In Fig.~\ref{fig:fig2}(a), the energy expectation values $E \equiv \langle {\cal H} \rangle$ for the several states 
in $(\Delta/t_{\rm A}, J/t_{\rm A})=(12.2, 4)$ are plotted as functions of $x$. 
The transfer integrals are chosen to be $t_B/t_A=1/4$. 
To show the numerical data clearly, we plot $ E'=(E/t_A)+C x$ with a numerical constant $C$, instead of $E$. 
This transformation does not affect the Maxwell's construction introduced below. 
The results in the AFM HS-ML are not plotted, because of their higher energy values than others.  
We also present, in Fig.~\ref{fig:fig3}, 
a ratio of the LS sites to the LS and HS ones in the mixed states defined by $R_{LS}=n_{LS}/(n_{LS}+n_{HS})$.  
Here $n_{LS}$ ($n_{HS}$) is a number of the sites where the LS (HS) state is realized. 
As shown in Fig.~\ref{fig:fig2}(a), the LS state, where holes are doped into the B band, is destabilized monotonically with increasing $x$. 
On the other side, in a region of $x>0.5$, the FM HS-ML state is realized. 
In between the two regions, the mixed state is the lowest energy state. 
The mixed state is smoothly connected to the LS and HS ones in the low and high $x$ regions, respectively. 
As shown in Fig.~\ref{fig:fig3}, a discontinuous jump in the mixed state   
is seen around $x=0.25$; the system is changed from the LS dominant mixed state into the HS dominant one with $x$.  
It is noticeable that 
the $E'$ versus $x$ curve in the mixed state is convex 
in the region of $0<x<0.33$.
That is, by following the Maxwell's construction, 
the PS of the LS-BI and the FM HS dominant mixed states is more stabilized 
than the homogeneous phase in this region of $x$. 
In the Fig.~\ref{fig:fig2}(b), we show the results in $(\Delta/t_{\rm A}, J/t_{\rm A})=(8.25, 2.5)$ 
where the system at $x=0$ is close to the ML phase [see Fig.~\ref{fig:fig1}(a)]. 
The PS appears, but its region is shrunken. 

The magnetization per site in the lowest energy state 
defined by $M(x)=(1/N)\langle \sum_i S^z_i \rangle$ is plotted in Fig.~\ref{fig:fig3}. 
A zero magnetization at $x=0$ reflects the LS-BI ground state. 
In a high doped region of $x>0.33$, the magnetization data almost follow a relation   
$M(x) \simeq (1+x)/2$: the system is expected to consist of the $N/2$ HS sites, the $(1/2-x)N$ LS ones, and the $xN$ singly electron occupied ones. 
In this scheme, we obtain $R_{LS}=(1-2x)/(2-2x)$ which is consistent with the numerical data of $R_{LS}$ in $x>0.33$. 
Between $x=0$ and $0.33$, where the PS is realized, 
$M(x=0)$ and $M(x=0.33)$ are connected by a straight line according to the volume-fraction rule in the Maxwell's construction. 
The slope of $M(x)$ is about three times higher than 
$M(x)=x/2$ which is expected in the hole doping into the LS-BI phase. 
This is qualitatively consistent with the experimental observations in the magnetization where doped holes induce high spin value.~\cite{yamaguchi,okamoto} 

\begin{figure}[t]
\begin{center}
\includegraphics[width=0.8\columnwidth,clip]{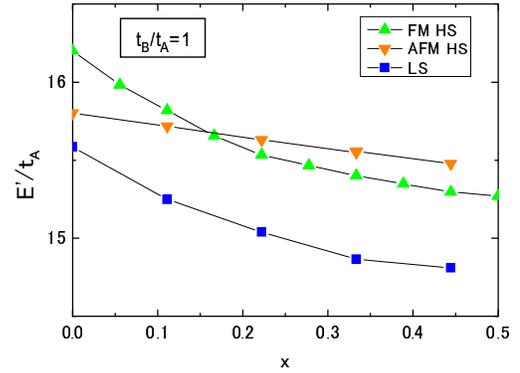}
\caption{(color online) 
Hole concentration dependences of the energy expectations for several states 
where the electron transfer integrals are chosen to be equal as $t_B/t_A=1$. 
Other parameters are taken to be 
$(\Delta/t_{\rm A}, J/t_{\rm A})=(12.2, 4)$, 
and a constant parameter $C$ in the definition of $E'$ is taken to be 8. 
}
\label{fig:fig4}
\end{center}
\end{figure}

We now address an origin of the electronic PS where the spin-state 
degree of freedom is concerned. 
In Fig.~\ref{fig:fig4}, we present the hole concentration dependence of the energy expectations where 
the band widths are set to be equal with each other, $t_B/t_A=1$. 
As well as the calculation in Fig.~\ref{fig:fig2}(a),
the energy parameters are taken to be $(\Delta/t_{\rm A}, J/t_{\rm A})=(12.2, 4)$ which 
is close to the LS-HS phase boundary at $x=0$ [see Fig.~\ref{fig:fig1}(b)]. 
The mixed state is not obtained in the simulation. 
In all region of $x$ up to $x=0.45$, the LS state is the lowest ground state, and 
neither the spin-state transition nor the PS occur. 
%Although the calculated data shown here is limited in one parameter set, we confirm that this tendency, the equal band widths suppresses the spin-state transition and PS, is general in the present calculation. 
The difference of the band widths in the two orbitals is of essence in the electronic PS phenomena. 

\begin{figure}[t]
\begin{center}
\includegraphics[width=0.8\columnwidth,clip]{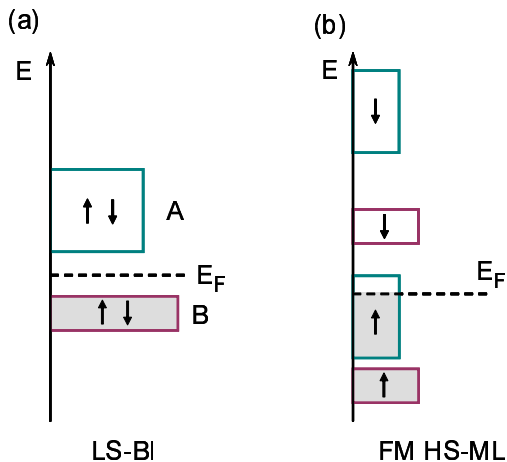}
\caption{(color online) 
Schematic density-of-states in the LS-BI state at $x=0$ 
and that in the HS-ML one in a high hole doped region. 
}
\label{fig:fig5}
\end{center}
\end{figure}
To clarify the mechanism of PS furthermore, 
schematic pictures of the density of states (DOS) in the LS-BI at $x=0$ and the FM HS-ML 
in a high hole doped region are 
presented in Fig.~\ref{fig:fig5}. 
For simplicity, detailed shapes of DOS are not taken into account. 
In LS-BI state at $x=0$, the fermi level is located inside of the band gap between 
the A and B bands. 
The band width in the A band is larger than that in B. 
On the other hand, in the FM HS-ML state which is realized in 
$x \gtrsim 0.5$ in Fig.~\ref{fig:fig2}(a), 
the system is a doped MI with ferromagnetic spin polarization. 
The fermi level is located in the A band. 
Because of the large band width in the A  band, there is a large kinetic energy gain in comparison 
with the doped LS-BI state where the fermi level is located in the B band in the rigid band scheme. 
%That is, the center of mass for the electron occupied state is lower in the FM HS-ML state that that in the doped LS-BI one. 
This kinetic energy gain is the origin of the spin state transition by doping. 
It is shown in Fig.~\ref{fig:fig4} that, when the equal band widths are assumed, 
the $E'$ v.s. $x$ curves for the LS-ML and FM HS-ML states are almost parallel and 
do not cross with each other. 
This data implies that there is no difference in the kinetic energy gains for the two states,
when the band widths are assumed to be equal. 
The present PS phenomena are also attributed to this band width difference as follows. 
In the rigid-band sense, by doping of holes in the LS-BI state, 
the fermi level falls into the top of the B band from the middle of the gap in Fig.~\ref{fig:fig5}(a). 
If we suppose that this state is realized in a low $x$ region and is transferred into the FM HS-ML state shown in Fig.~\ref{fig:fig5}(b) with increasing $x$, 
the fermi level is increased with increasing hole concentration 
because of the different band widths. 
This is nothing but the negative charge compressibility $\kappa=(\partial \mu /\partial x)<0$ with the chemical potential $\mu$, i.e. appearance of the electronic PS.

Finally, we discuss implications of the perovskite cobaltites. 
The obtained PS between the insulating nonmagnetic state and the hole-rich FM one 
is qualitatively consistent with the inhomogeneity suggested by a number of experiments. 
The PS and the spin-state transition are attributed to the band-width difference of the two bands corresponding to the $e_g$ and $t_{2g}$ bands in the perovskite cobaltites. 
This electronic PS is robust by changing the model parameter values, except for $t_B/t_A$, when the non-doped 
system is located near the phase boundary between the LS-BI and HS-MI. 
The present phenomena are different from the previous PS's discussed in the high-Tc cuprates and the manganites where 
the long-range spin/orbital orders are realized in the MI's;  
the spatial segregations occur between the long-range ordered MI and the ML states where 
the superexchange interaction energy and the kinetic one of doped holes are 
separately gained in the different spatial regions. 
Our scenario of the PS based on the band-width difference may be checked experimentally by 
adjusting the tolerance factor, i.e. the Co-O-Co bond angle; the smaller tolerance factor implies 
the smaller (larger) band width in the $e_g$ $(t_{2g})$ orbitals, and suppression of the PS.  
Detailed values of $x$ where the PS is realized, and a typical size of the clusters remain 
as questions. 
Several factors not considered here, the intermediate-spin state, the long-range Coulomb interaction, 
the lattice volume depending on the spin states, and so on, are required to answer these questions. 

\par
Authors would like to thank H.~Yokoyama and H.~Takashima for their valuable discussions. 
This work was supported by JSPS KAKENHI, TOKUTEI from MEXT, and Grand challenges in next-generation integrated nanoscience. 
$^{\dagger}$ Present address: The bank of Tokyo-Mitsubishi UFJ, Tokyo, Japan. \\
$^{\ast}$ Present address: Chiba Institute of Technology, Tsudanuma, Chiba 275-0016, Japan.


\begin{references}
\bibitem{variation}
The 10 inequivalent configurations considered here are 
$|0 \rangle$, 
$|A^\sigma \rangle$,
$|B^\sigma \rangle$,
$|A^\sigma B^{\bar \sigma} \rangle$,
$|A^\sigma B^\sigma \rangle$,
$|B^\uparrow B^\downarrow \rangle$,
$|A^\uparrow A^\downarrow \rangle$,
$|A^\sigma B^\uparrow B^\downarrow \rangle$,
$|A^\uparrow A^\downarrow B^\sigma \rangle$,
$|A^\uparrow A^\downarrow B^\uparrow B^\downarrow \rangle$ 
with $\sigma=\uparrow$ and $\downarrow$, 
where $A^\sigma (B^\sigma)$ impies that 
the A (B) orbital is occupied by the up (down) spin electron.
\end{references}
\end{document}